\documentclass[aps,prd,twocolumn,showpacs,floatfix,superscriptaddress,
preprintnumbers,nofootinbib]{revtex4-1}

\usepackage{graphicx,amssymb,amsmath,amsfonts,color}
\usepackage{subcaption}

\newcommand{\gcmb}{\gamma_{\textsc{cmb}}}

\newcommand{\gcmbebl}{\gamma_{\textsc{cmb},\textsc{ebl}}} %

\begin{document} %
	\title{Cascade photons as test of protons in UHECR} %
	
	\author{V.~Berezinsky}
	\affiliation{INFN, Laboratori Nazionali del Gran Sasso, Assergi
		(AQ), 67010, Italy}
	\affiliation{INFN, Gran Sasso
		Science Institute, viale F.Crispi 7, 67100 L'Aquila, Italy} %
	
	\author{A.~Gazizov}
	\affiliation{INFN, Laboratori Nazionali del Gran Sasso, Assergi (AQ),
		67010, Italy} %
	
	\author{O.~Kalashev}
	\affiliation{Institute for Nuclear Research of the Russian Academy of
		Sciences, Moscow 117312, Russia}
	\email{kalashev@inr.ac.ru}
	
	\date{\today} %
	
	\begin{abstract} %
		An isotropic component of high energy $\gamma$-ray spectrum measured
		by Fermi LAT constrains the proton component of UHECR. The strongest
		restriction comes from the highest, $(580-820)$~GeV, energy bin. One
		more constraint on the proton component is provided by the IceCube
		upper bound on ultrahigh energy cosmogenic neutrino flux. We study the
		influence of these restrictions on the source properties, such as
		evolution and distribution of sources, their energy spectrum and
		admixture of nuclei. We also study the sensitivity of restrictions to
		various Fermi LAT galactic foreground models (model B being less
		restrictive), to the choice of extragalactic background light model
		and to overall normalization of the energy spectrum. We claim that the
		$\gamma$-ray-cascade constraints are stronger than the neutrino ones,
		and that however many proton models are viable. The basic parameters
		of such models are relatively large $\gamma_g$ and not very large
		$z_{\max}$. The allowance for H$e^4$ admixture also relaxes the
		restrictions. However we foresee that future CTA measurements of
		$\gamma$-ray spectrum at $E_\gamma \simeq (600 - 800)$~GeV, as well as
		resolving of more individual $\gamma$-ray sources, may rule out the
		proton-dominated cosmic ray models.
	\end{abstract} %
 
\pacs{98.70.Sa, 95.85.Ry cosmic rays} %

\maketitle %

\section{Introduction} %
\label{sec:intro} %
In spite of tremendous technical progress, the long-standing crisis in
ultrahigh-energy cosmic rays (UHECR) is not yet settled. In short, the
essence of this crisis is the difference in mass compositions at
energies above $4$ EeV obtained by Pierre Auger Observatory (PAO)
\cite{Aab:2014aea} on the one hand, and HiRes \cite{Fedorova:2007cta}
(closed in 2009) and Telescope Array (TA) \cite{Abbasi:2014sfa} on the
other hand. PAO shows a steadily heavier mass composition with energy
increasing starting from protons (or protons and Helium) at $E \simeq
1$ EeV up to heavier nuclei at $E \simeq 30$~EeV. The two other biggest
experiments are consistent with a pure proton composition, or maybe
with mixed proton + Helium one, at all energies $E \gtrsim 1$~EeV. It
should be noted that all three experiments use the same fluorescent
light detection technique for measurement of the mass composition.

In this situation indirect methods of the mass composition detection
become important, though they cannot dismiss the necessity of direct
measurements. In fact, the energy spectrum is strongly involved in the
conflict with mass composition. 

The pure proton composition leaves remarkable features in the energy
spectrum. Propagating through CMB, UHE protons undergo photopion
production 
\begin{equation} %
\label{pion-prod} %
p + \gcmb \to \pi^{\pm,0} + X, %
\end{equation} %
and pair-production 
\begin{equation} %
\label{pair-prod} %
p + \gcmb \to e^+ + e^- + p, %
\end{equation} %
which result in a sharp steepening of the spectrum due to
Eq.~(\ref{pion-prod}) at highest energies, called
Greisen-Zatsepin-Kuzmin (GZK) cutoff \cite{Greisen:1966jv,
Zatsepin:1966jv}, and in a shallow deepening of the spectrum at lower
energies owing to (\ref{pair-prod}) called {\em dip}
\cite{Blumenthal:1970nn}.

Presence of these features in the energy spectrum proves the 
{\em proton composition} of UHECR.

All three aforementioned detectors have observed the sharp steepening
in the end of the energy spectrum and all three collaborations claim
that the observed steepening is consistent with GZK cutoff. But while
the steepenings observed by HiRes and TA before 2015 agree well with
the theoretically predicted value $E\approx 50$~EeV, namely $E = 56.2
\pm 5.1$~EeV for HiRes and $E = 48 \pm 1.0$~EeV for TA, the cutoff
energy in the case of PAO is noticeably lower, $E =
25.7^{+1.1}_{-1.2}$~EeV. For the PAO collaboration, this is an
expected result since PAO observes a non-proton mass composition at
highest energies.

Before 2015 both HiRes and TA also observed the GZK cutoff in the
integral spectrum at $E_{1/2} = 53.7 \pm 8$~ EeV, in a good
agreement with the theoretical prediction $E_{1/2} \approx 52.5$~ EeV
\cite{Berezinsky:1988wi}. However, in new higher statistics data
\cite{Kalashev:2014xna} the GZK cutoff is not seen as clearly as
before both in differential and in integral spectra.

For the last 20 years, the scientific community was hypnotized by the
GZK cutoff, more precisely by its absence. And the glory of this
phenomenon left in shadow another feature, the dip, which is also a
signature of protons interacting with CMB photons. This feature is
quite faint, but it is located at lower, $(1 - 40)$~EeV, energies where
statistics is much higher than in the case of GZK cutoff. 

Calculated for the ordinary UHECR spectrum, the dip is a
model-dependent feature. Its shape depends on many phenomena and
parameters, such as the way of propagation (rectilinear or diffusive),
index of the generation spectrum $\gamma_g$, parameters of
cosmological evolution and especially the mass composition. But in
terms of {\em modification factor} $\eta(E)$ \cite{Berezinsky:2002nc,
Aloisio:2006wv}, dip becomes considerably less model-dependent, still
remaining different for protons and nuclei.

Defined for protons, the modification factor is a ratio of proton
spectrum $J_p(E)$, calculated with all energy losses included, and
of so-called unmodified spectrum, $J_{\rm unm}(E)$, which accounts
only for adiabatic (due to red-shift) energy loss:
\begin{equation} %
\label{modf}
\eta(E) = J_p(E)/J_{\rm unm}(E). 
\end{equation} %

Modification factor is an excellent tool for {\em interaction
signatures} \cite{Berezinsky:2005cq}. According to (\ref{modf}),
interactions (\ref{pion-prod}) and (\ref{pair-prod}) enter only the
numerator remaining unsuppressed in $\eta(E)$, while most other
phenomena entering both the numerator and the denominator are either
suppressed or even cancelled in the modification factor.

This property is especially pronounced for the dip modification
factor. According to Ref.~\cite{Berezinsky:2002nc, Aloisio:2006wv},
the {\em theoretical} dip modification factors depend very weakly on
generation index $\gamma_g$, $E_{\max}$ and on such characteristics as
propagation mode, average source separation, local source overdensity
or deficit etc. Calculated for different $\gamma_g= 2.0$ and $2.7$
(see Fig.~2 of Ref.~\cite{Berezinsky:2002nc} and Fig.~3 of
Ref.~\cite{Aloisio:2012ba}), they are practically indistinguishable at
all energies.

While cosmological evolution of sources just moderately modifies
$\eta(E)$, an admixture of nuclei changes it significantly. Therefore
one can define the {\em dip model} in terms of modification factor as
one strongly dominated by protons and with weak cosmological evolution
(see Ref.~\cite{Berezinsky:2002nc} where the evolution is taken as
that of AGN in X-ray observations).

Above, the theoretical modification factor was discussed. The {\em
observational} modification factor is given by the ratio of the
observed flux $J_{\rm obs}(E)$ and unmodified spectrum $J_{\rm unm}(E)
\propto E^{-\gamma_g}$. Defined up to normalization constant,
\begin{equation} %
\label{eta-obs}
\eta_{\rm obs}(E) \propto J_{\rm obs}(E)/E^{-\gamma_g}. %
\end{equation} %
Here $\gamma_g$ is an exponent of the proton generation function
$Q_{\rm gen}(E_g) \propto E_g^{-\gamma_g}$ in terms of initial proton
energies $E_g$.

To fit $\eta(E)$ to $\eta_{\rm obs}(E)$ one has just two free
parameters, $\gamma_g$ and the overall normalization factor for
$\gtrsim 20$ energy bins of each experiment. The comparison is shown
in Fig.~8 of Ref.~\cite{Berezinsky:2002nc} and Fig.~4 of
Ref.~\cite{Aloisio:2012ba}. It is clear that both the proton
pair-production dip and the beginning of GZK cutoff up to $80$~EeV
were well confirmed by data of Akeno-AGASA, HiRes, Yakutsk and
Telescope Array with data of 2013. And only two free parameters were
needed for the description of $20 - 30$ energy bins in each of four
experiments. The values of $\gamma_g$ providing this agreement were
fixed as $2.6 - 2.7$.

The ankle observed in all these experiments is well described as dip
produced in the pair-production process (\ref{pair-prod}) and not by
transition from galactic to extragalactic cosmic rays. So, an
excellent confirmation of dip and GZK cutoff in terms of modification
factor in all abovementioned experiments evidences for the proton
composition of UHECR \cite{Aloisio:2012ba}. 

Another indirect method to distinguish proton-dominated cosmic ray
models from nuclei-dominated ones is given by measurements of {\em
cosmogenic neutrino} flux. Protons are much more efficient in the
production of neutrinos than nuclei. Being first proposed in 1969
\cite{Beresinsky:1969qj}, cosmogenic neutrinos and their production
have been studied in many works \cite{Engel:2001hd, Kalashev:2002kx,
Decerprit:2011qe, Desiati:2012df}. The observational upper limits on
cosmogenic neutrinos have been recently obtained at $E > 10^{16}$~eV
in IceCube detector~\cite{Ishihara:2015xuv,Aartsen:2016ngq} and at $E > 10^{17}$~eV in
PAO \cite{Abraham:2007rj}. These upper limits constrain only some
extreme models of UHECR with hard injection spectrum, strong evolution
and relatively high $z_{\rm max}$~\cite{Heinze:2015hhp}. Generally,
they do not dismiss both proton and nuclei models.

In the present work we study the allowed class of proton UHECR models,
which fit the observed UHECR spectrum. We also calculate the diffuse
neutrino fluxes for these models. In some cases the produced
neutrino flux exceeds the upper limit of IceCube; these models are
qualified as excluded. But the majority of models analysed in this
paper are allowed by the IceCube upper limit.

And finally, the most severe indirect constraint on proton models is
imposed by the observed diffuse $\gamma$-radiation. We consider this
restriction below, first discussing a specific problem of cosmogenic
neutrino and diffuse $\gamma$-ray flux. The neutrino flux is directly
connected with the diffuse $\gamma$-radiation in case it has
electro-magnetic (EM) cascade nature. The flux of cosmogenic neutrinos
is strongly suppressed if the flux of cascade $\gamma$-radiation is
low. This phenomenon was first noticed in
Ref.~\cite{Berezinsky:1975zz}, where a relation between cosmogenic
neutrino flux $J_{\nu}(E)$ and energy density $\omega_{\rm cas}$ of
cascade $\gamma$-radiation was obtained as
\begin{equation} %
\label{nu-limit} %
E^2 J_{\nu}(E) < \frac{c}{4\pi}\, \omega_{\rm cas}. %
\end{equation} %
In Ref.~\cite{Berezinsky:2010xa} this formalism was further developed.
Using the Fermi EGRB $\gamma$-ray flux, it was found that maximum
allowed EM cascade energy density $\omega_{\rm cas} = 5.8 \times
10^{-7}$~eV/cm$^3$. The cosmogenic neutrino flux was found to be below
the IceCube upper limit \cite{Berezinsky:2010xa}, but still detectable
in near future~\cite{Ahlers:2010fw, Gelmini:2011kg}.

The properties of extragalactic diffuse $\gamma$-radiation render a
powerful tool for distinguishing between proton or nuclei dominances in
the UHECR spectrum. This information can be obtained from observation
of the energy spectrum of diffuse $\gamma$-radiation and its
$\omega_{\rm cas}$. A historical development of observations tends to
the diminishing of the role of protons as a source of the observed
extragalactic diffuse $\gamma$-radiation.

The research approach has been initiated by the study of diffuse
galactic $\gamma$-radiation on SAS-2 satellite. In 1975 it has
demonstrated \cite{Fichtel:1975aq} that this radiation is produced by
galactic cosmic rays. In the 1990s the EGRET detector on board of
Compton Gamma Ray Observatory measured extragalactic diffuse
$\gamma$-ray emission in energy interval $\rm 30\, MeV - 100\, GeV$
\cite{Sreekumar:1997un} and detected the extragalactic $\gamma$-ray
sources, including blazars.

In Refs.~\cite{Berezinsky:1975zz, Berezinsky:1991aa} the EGRET data
were used to put upper limits on diffuse fluxes of UHECR and
cosmogenic neutrinos. The observed diffuse $\gamma$-radiation in these
works was interpreted as $\gamma$-radiation from EM cascades initiated
by protons. The cascade energy spectrum was estimated as $\propto
E^{-2}$ \cite{Berezinsky:1975zz}, in agreement with measured in
Ref.~\cite{Sreekumar:1997un} gamma-ray spectrum $\propto E^{-\gamma}$
with $\gamma = 2.10 \pm 0.03$. The physical quantity which
characterizes UHE neutrinos and CR diffuse flux was given in these
calculations by energy density of the cascade radiation $\omega_{\rm
cas}$. Using the EGRET data, it as found in
Refs.~\cite{Berezinsky:1975zz, Berezinsky:1991aa} that $\omega_{\rm
cas} = 5 \times 10^{-6}$~eV/cm$^3$.

The first $10$ months of Fermi-LAT observations \cite{Abdo:2010nz}
have put a stronger limit on the isotropic diffuse gamma-ray
background (IGRB) in energy interval $\rm 200\, MeV - 120\, GeV$. A
more steep index of the power-law spectrum, $\gamma= 2.41 \pm 0.05$,
was found. The analysis of Ref.~\cite{Berezinsky:2010xa} gave lower
IGRB and, respectively, lower cascade energy density $\omega_{\rm
cas}= 5.8\times 10^{-7}$~eV/cm$^3$.

Regular lowering of the $\omega_{\rm cas}$ since the first SAS-2
satellite measurement means the diminishing of $\gamma$-ray cascade
flux and hence of associated with it UHE proton flux. Analysis
\cite{Berezinsky:2010xa} of the data including $\omega_{\rm cas}$ in
terms of UHECR proton models demonstrates that many proton models
survive, though some of them, mostly those with strong cosmological
evolution, are excluded.

More stringent limit on proton component of UHECR can be extracted
from 50 months observation of Fermi LAT \cite{Ackermann:2014usa}. The
limit becomes stronger due to Highest Energy Bin (HEB) at $(580 -
820)$~GeV, where the Fermi-LAT flux is particularly low. This effect
is analysed in Refs.~\cite{Liu:2016brs, Gavish:2016tfl,
Berezinsky:2016feh} with rather an extreme conclusion in
Ref.~\cite{Liu:2016brs}. Here we argue that using reasonable galactic
foreground in Fermi LAT analysis (model B) and extragalactic
background light (EBL) model we keep the proton models alive.

Nevertheless, we admit that isotropic $\gamma$-radiation looks like a
serious potential problem for models with proton composition. The
difference in generation indexes of predicted cascade gamma-ray
spectrum $\gamma = 1.9$ and observed by Fermi LAT, $\gamma = 2.4$ shows
that proton-induced composition of $\gamma$-rays must be much smaller
than one observed. There easily could be more unresolved sources in
IGRB flux. On the other side, the clearly seen dip and GZK cutoff in
modification factor analyses strongly support the proton composition. 
We expect that future CTA \cite{Consortium:2010bc} data may radically 
change the situation.

\section{UHECR and cascade radiation} %
\label{sec:general} %
In this section we first construct (subsection~\ref{sec:models}) the
standard proton models to fit the observed spectra of UHECR and
calculate the produced fluxes of the cascade EM radiation. Protons
with energies $(2 - 20)$~EeV at present create $e^+e^-$ pairs on CMB
photons (\ref{pair-prod}). The produced electrons and positrons
initiate EM cascades interacting with EBL and CMB target photons. The
spectra of cascading photons are calculated using two methods: by
solving kinetic equations and performing MC simulations. The isotropic
diffuse $\gamma$-ray background measurements by Fermi
LAT~\cite{Ackermann:2014usa} put the upper limit on the flux of
cascade photons. The constraint is especially strong at the highest
energy bin, i.e.\ at $(580 - 800)$ GeV.

The IGRB flux is strongly model-dependent especially at HEB since it
is derived by subtraction of the simulated galactic contribution from
observational data. We take into account the model dependence by
considering all three galactic foreground models originally used in
IGRB calculations. The model B leads to highest IGRB estimate and
therefore is the least restrictive. We also account for uncertainties
in EBL by using two EBL models of Refs.~\cite{Kneiske:2003tx} and
\cite{Inoue:2012bk}. As a result, we obtained two tables of main
proton source model characteristics, which correspond to two EBL
models with different status of the agreement with IGRB flux. The
further increase of the number of models which respect the HEB
restriction in the two aforementioned tables, may be reached by taking
into account the possible systematic errors in measured energies of
UHECR. Shifting the whole energy spectrum downwards results in
decreasing of $e^+e^-$ production rate, so that more models become
allowed.

In subsection~\ref{sec:1-4EeV}, following Ref.~\cite{Liu:2016brs}, we
address a question whether the proton component observed at $(1 -
4)$~EeV in all experiments (HiRes, TA, and Auger) contradicts to Fermi
LAT IGRB flux. For this purpose we construct auxiliary models with
cutoffs at high and low energies to imitate spectrum in the discussed
$(1 - 4)$~EeV energy range. We find that in many cases the observed
proton flux at $(1 - 4)$~EeV is allowed by the Fermi LAT IGRB. A
contradiction with Ref.~\cite{Liu:2016brs} is mainly explained by
using of model B for galactic contribution in the Fermi LAT
experiment.

In subsection~\ref{sec:helium} we analyze the proton component with
nuclei admixture in the form of Helium. The basic idea behind this
proposal is that experimentally He$^4$ is difficult to tell from
protons. Helium is less efficient in the production of $e^+e^-$ pairs
and thus the cascade flux is suppressed. In fact, the situation is
more complicated and different for hard and soft generation spectra.
For small generation indexes, e.g.\ $\gamma_g = 2.1$ the secondary
proton component from Helium photo-disintegration is comparable with
the primary proton flux and thus the $\gamma$-ray component is
suppressed but a little. For large generation indexes, e.g.\ $\gamma_g
= 2.6$, the secondary protons from Helium decay are strongly
suppressed and thus Helium-produced flux of photons is small. However,
extra component is required in this case to fit the UHECR observations
above $4$~EeV.

In subsection~\ref{sec:sources} we calculate the red-shift
distribution of points of cascade $\gamma$-ray production and redshift
distribution of parent protons in the models of $p\gamma$-production
of the primaries for cascade photons.

\subsection{Standard proton models} %
\label{sec:models} %
We consider simple phenomenological models of homogeneously
distributed sources emitting ultrahigh energy protons with
power-law generation spectra 
\begin{equation} %
\label{Q_p} %
Q_p(E,z) \propto n(z) \left(\frac{E}{E_0}\right)^{-\gamma_g}, \quad
E \in [E_{\min},  E_{\max}], %
\end{equation} %
where $E_0$ is an arbitrary normalization energy. Below unless we
state explicitly, we cut the injection spectrum below $E_{\min}=0.1$
EeV and above $E_{\max} = 10^{2.5}$ EeV without loss of generality.
Indeed the main contribution to the EM cascade comes from protons with
energies in the interval from $1$~EeV to a few EeV unless the
injection spectrum is too flat ($\gamma_g \leq 2$) which is forbidden
anyway as it will be demonstrated below. Also note, that
models with $E_{\max}>10^{2.5}$ EeV don't substantially improve UHECR
fit but may overproduce secondary $\nu$-flux. We also introduce the
evolution of source density with redshift $z$ given by the term $n(z)$
assuming however that the source spectrum shape does not depend on
$z$. For the evolution term we use a general form

\begin{equation} %
\label{evol} %
n(z)=n(0)(1+z)^{3+m} \quad \mbox{for } 0 \leq z \leq z_{\max}, %
\end{equation} %
where the case of $m=0$ corresponds to the constant comoving source
density. We also consider a specific case of source density
proportional to the star formation rate (SFR)~\cite{Yuksel:2008cu}:
\begin{equation} %
n_{\rm SFR}(z) \propto (1+z)^3\left \{ %
\begin{array}{lll} %
(1+z)^{3.4}, \;\; & z \leq 1 \\ %
(1+z)^{-0.3}, \;\; & 1 < z \leq 4\\ %
(1+z)^{-3.5}, \;\; &z>4. 
\end{array} %
\right. %
\end{equation} %
With these assumptions and with $\gamma_g$ varying in a wide range $2
- 2.7$, we include in consideration a large class of models used in
literature.

During propagation to an observer, protons lose their energy through pion
(\ref{pion-prod}) and $e^+e^-$-pair production (\ref{pair-prod}) on
CMB and EBL. Both processes give rise to EM cascades since secondary
electrons and photons are produced with energies above the threshold
for $e^+e^-$-pair production. High energy photons are produced by
inverse Compton $e^\pm$ scattering off CMB photons
\begin{equation} %
\label{IC} %
e^\pm + \gcmb \to e^\pm + \gamma, %
\end{equation} %
and new high energy $e^\pm$ pairs arise in  
\begin{equation} %
\label{gg-pair} %
\gamma + \gcmbebl \to e^+ + e^- %
\end{equation} %
collisions of $\gamma$-rays with CMB and EBL.

This chain of reactions proceeds until photons become sterile forming
the diffuse $\gamma$-ray background, and the latter can be compared
with the measurements of Fermi LAT~\cite{Ackermann:2014usa}. We
simulate propagation of protons and development of secondary EM
cascades using the code of Refs.~\cite{Kalashev:2014xna,
Gelmini:2011kg} that allows solving transport equations in
\mbox{1-D}. The solutions are double-checked by an independent Monte
Carlo code previously used in Ref.~\cite{Berezinsky:2016feh}.

Although energy density of EBL is $\sim 15$ times lower than that of
CMB, the EBL photons play a crucial role in attenuation of EM cascades
below the pair production threshold on CMB, i.e.\ at $E_{\rm \gamma}
\lesssim 100 $~TeV. The EBL energy spectrum has a characteristic
two-bump shape with a near-infrared bump at $\sim 1$~eV produced by
direct starlight emission and a far-infrared bump at the energy
around $0.01$~eV produced by starlight scattering off dust. Direct
measurements provide just an upper bound on the EBL intensity because
of a much stronger foreground of zodiacal light from the Solar system;
the latter is to be subtracted from observations. The lower bounds on
the EBL intensity may be estimated using source counts in deep
observations by infrared and optical telescopes.

The present upper and lower bounds on EBL at different wavelengths are
summarized in Fig.~7 of Ref.~\cite{Dwek:2012nb}. In the literature,
there are additional constraints of EBL based on observations of
distant blazars. These bounds were derived using the attenuation of
photons without an account for the possible contribution of secondary
$\gamma$-ray signal from protons; it means that these constraints may
be relaxed~\cite{Essey:2009ju}. The limits based on
GRBs~\cite{Abdo:2010kz} observations remain unaffected, however, they
are only applicable to the highest energy part of EBL spectrum.

In this work we use one of the latest EBL model by Inoue et
al.~\cite{Inoue:2012bk} which is close to the minimal estimate. In
parallel we use the popular model of Kneiske et
al.~\cite{Kneiske:2003tx} that provides a larger EBL density. It will
be shown that the choice of EBL model has a noticeable effect on the
cascade flux.

Among the cosmic ray models we look for those providing reasonable
fits to the Telescope Array~\cite{TheTelescopeArray:2015mgw} and
HiRes~\cite{Abbasi:2007sv} spectra. The energy spectra of both
detectors are in a good agreement, and both observe the mass
composition compatible with pure protons.

It is important to note that we do not require the perfect fitting of
the cosmic ray data, especially in terms of best $\chi^2$. The reasons
are as follows:
\begin{itemize} %
\item %
at present, the systematic errors in all UHECR experiments are much
larger than the statistical ones, %
\item %
demanding the best fit can be too limiting for the {\em exclusion} of
some hypotheses, %
\item %
we use just simple phenomenological source models, e.g.\ assuming
their homogeneous distribution, %
\item %
we consider pure proton models, while an admixture of nuclei is quite
possible; this is especially true for Helium nuclei, which are
difficult to distinguish experimentally from protons. %
\end{itemize} %

With the above assumptions, we are able to construct reasonable fits
to data with sufficiently high $E_{\rm max}$ and power-law indexes in
the range $2 \leq \gamma_g \leq 2.7$, and to choose appropriate
evolution parameter $m$ in Eq.~(\ref{evol}) in the range $0 \leq m
\leq 7$. To fit UHECR data at $E \sim (1-10)$~EeV , the harder
injection spectra require the stronger evolution (for illustration see
e.g.\ Fig.~4a of Ref.~\cite{Gelmini:2011kg}). The choice of $z_{\max}$
has a weak effect on UHECR spectrum, provided that $z_{\max} \gtrsim
1$, but it does have the effect on fluxes of secondary $\gamma$-rays
and neutrinos produced in this model.

We should also note that hard injection spectra in combination with
strong evolution bring to a proton flux exceeding the KASCADE-Grande
measurements \cite{Arteaga-Velazquez:2015jpx}. We do not deny such
models since the average extragalactic injection spectrum may have a
broken power-law form. It may be also suppressed even stronger below
$1$~EeV, for example, because of sources distribution in maximal
energies~\cite{Kachelriess:2005xh}. Otherwise, to avoid proton
overproduction at $E \lesssim 1$~EeV, one should assume lower maximum
red-shifts, $z_{\max} \lesssim 0.7$.

The Fermi LAT collaboration has presented the measurements of
isotropic gamma-ray background, IGRB, and of the total extragalactic
gamma-ray background (EGB), the latter being composed of IGRB and the
flux from resolved extragalactic sources~\cite{Ackermann:2014usa}. The
EGB flux is in turn calculated by subtraction of galactic foreground
from observational data.

We normalize proton models by fitting the TA spectrum and compare the
calculated integral fluxes of the cascade $\gamma$-radiation,
$\Phi^{cas}_i$, with the IGRB integral flux $\Phi^{\rm IGRB}_i$ in
each energy bin of the Fermi LAT diffuse background data. Namely, for
each bin $i$ we calculate ratios %
\[ %
\eta_i = \Phi_i^{\rm cas}/\Phi_i^{\rm IGRB}. %
\] %
These ratios have maximum in some bin $i'$ and we introduce the value
$\eta_\gamma$ as
\begin{equation} %
\label{ratio} %
\eta_\gamma= \max (\Phi_i^{\rm cas}/\Phi_i^{\rm IGRB}). %
\end{equation} %
A strong criterion for the model consistency with data is given by
\begin{equation} %
\label{ratioIGRB} %
\eta_\gamma\leq 1. %
\end{equation} %
Most often (but not always !) the criterion of consistency takes
place in the highest energy bin, where $\Phi_i^{\rm IGRB}$ has
minimum.

An analysis of the recent Fermi LAT data \cite{DiMauro:2016cbj} shows
that a considerable fraction of EGB events with energies above
$50$~GeV may be attributed to unresolved $\gamma$-ray sources, mostly
to blazars. According to this analysis, the contribution of blazars to
EGB flux reaches $86^{+16}_{-14}\%$. This implies even stronger bound
on the true isotropic flux %
\begin{equation} %
\label{ratio50GeV} %
\tilde{\eta}_{\gamma}\equiv \frac{\int_{50 GeV}^{\infty} 
\Phi_{\rm \gamma}^{\rm cas}(E)dE }
{0.28\int_{50 GeV}^{\infty}\Phi_{\rm EGB}(E)dE}\leq 1.
\end{equation} %

The denominator of Eq.~(\ref{ratio50GeV}) stands for true isotropic
integral flux calculated under an assumption of minimal contribution
from all resolved and unresolved blazars. At $1\, \sigma$
approximation its fraction is just $0.86 - 0.14 = 0.72$ of the total
EGB flux. In this case, the fraction of isotropic component in EGB is
$1-0.72=0.28$, as it is included in the denominator of
Eq.~(\ref{ratio50GeV}).

In Tables~\ref{tableUnshifted} and~\ref{tableUnshiftedKneiskeBFebl}
the maximal fractions $\eta_{\gamma}$ and $\tilde{\eta}_{\gamma}$ are
shown for several representative models. The models with
$\eta_{\gamma} > 1$ or $\tilde{\eta}_{\gamma} > 1$ are in
contradiction with the Fermi LAT data. In this case the
$\min(1/\eta_{\gamma}, 1/\tilde{\eta}_{\gamma})$ could be roughly
interpreted as a maximal allowed fraction of protons in the source
spectrum. To be conservative, for the Fermi LAT IGRB and EGB fluxes we
use upper limits on the flux allowed by statistical and instrumental
errors. We also take into account the uncertainties in galactic 
foreground by considering all three models, A, B and C, used for
derivation of IGRB in Ref.~\cite{Ackermann:2014usa}.

 In addition, we calculate all-flavor flux of cosmogenic neutrinos, arising from decays of pions and
 neutrons produced in $p\gamma$-collisions, and the expectation value for the number of neutrino events $\bar{N_\nu}$ with energy $E_\nu>10$~PeV, where no events have been observed so far. For this we use the recently published IceCube exposure~\cite{Aartsen:2016ngq} for 7 years of observation and assume neutrino flavour ratio after propagation (1:1:1), which is roughly true for cosmogenic neutrinos. The values of $\bar{N_\nu}$ are shown in the last column of Table~\ref{tableUnshifted}. The models with $\bar{N_\nu}>2.3$ have Poisson likelihood less than 10\%.

\begin{table*}[t!] %
	\begin{center} %
		\begin{tabular}{c|c|c|c|c|c|c} \hline %
			$\gamma_g$  &  m  &  $z_{\rm max}$  &$\eta_{\rm\gamma}$
			($\tilde{\eta}_{\rm\gamma}$) [A] & $\eta_{\rm\gamma}$
			($\tilde{\eta}_{\rm\gamma}$) [B] & $\eta_{\rm\gamma}$
			($\tilde{\eta}_{\rm\gamma}$) [C] & $\bar{N_{\rm\nu}}$ \\ \hline
2.6 & 1 & 5 & 1.40 (0.59) & 0.94 (0.50) & 1.11 (0.57) & 0.78 \\
2.6 & 1 & 1 & 1.38 (0.46) & 0.93 (0.39) & 1.10 (0.44) & 0.31 \\
2.5 & 2 & 5 & 1.60 (0.87) & 1.07 (0.74) & 1.26 (0.84) & 2.24 \\
2.5 & 2 & 1 & 1.57 (0.60) & 1.05 (0.51) & 1.24 (0.58) & 0.48 \\
2.4 & SFR & 5 & 1.88 (1.20) & 1.26 (1.03) & 1.49 (1.16) & 2.28 \\
2.3 & 5 & 1 & 2.23 (1.38) & 1.49 (1.18) & 1.76 (1.33) & 1.72 \\
2.2 & 6 & 1 & 2.52 (1.86) & 1.69 (1.59) & 2.00 (1.79) & 2.88 \\
2.2 & 5 & 0.7 & 2.15 (0.83) & 1.44 (0.71) & 1.70 (0.80) & 0.99 \\
2.2 & 6 & 0.7 & 2.31 (0.99) & 1.55 (0.85) & 1.83 (0.95) & 1.19 \\
			\hline %
		\end{tabular} %
	\end{center} %
	\caption{
		Maximal ratios $\eta_{\rm \gamma}$,
		$\tilde{\eta}_{\rm\gamma}$ for galactic $\gamma$-ray foreground models
		A, B or C for several representative proton source models fitting TA spectrum. The ratios higher than 1 are in conflict with Fermi LAT data. Also shown the expectation value of the neutrino events $\bar{N_{\rm\nu}}$ with energy $E_\nu>10$~PeV assuming IceCube 7 year exposure from Fig.1 of Ref.~\cite{Aartsen:2016ngq}. Models with $\bar{N_{\rm\nu}}>2.3$ have Poisson probability less than 10\%. All spectra are calculated using the EBL model of Ref.~\cite{Inoue:2012bk}.
		} %
	\label{tableUnshifted} %
\end{table*} %

\begin{table*}[t!] %
\begin{center} %
\begin{tabular}{c|c|c|c|c|c|c} \hline %
$\gamma_g$  &  m  &  $z_{\rm max}$  &$\eta_{\rm\gamma}$
($\tilde{\eta}_{\rm\gamma}$) [A] & $\eta_{\rm\gamma}$
($\tilde{\eta}_{\rm\gamma}$) [B] & $\eta_{\rm\gamma}$
($\tilde{\eta}_{\rm\gamma}$) [C] & $\bar{N_{\rm\nu}}$ \\ \hline %
2.6 & 1 & 5 & 0.92 (0.66) & 0.61 (0.57) & 0.73 (0.64) & 0.78 \\ %
2.6 & 1 & 1 & 0.90 (0.48) & 0.60 (0.41) & 0.71 (0.47) & 0.31 \\ %
2.5 & 2 & 5 & 1.02 (1.03) & 0.68 (0.89) & 0.81 (1.00) & 2.24 \\ %
2.5 & 2 & 1 & 0.99 (0.63) & 0.66 (0.54) & 0.79 (0.61) & 0.48 \\ %
2.4 & SFR & 5 & 1.16 (1.34) & 0.78 (1.15) & 0.92 (1.30) & 2.28 \\ %
2.3 & 5 & 1 & 1.29 (1.47) & 0.87 (1.26) & 1.02 (1.42) & 1.72 \\ %
2.2 & 6 & 1 & 1.42 (2.00) & 0.95 (1.71) & 1.17 (1.93) & 2.88 \\ %
2.2 & 5 & 0.7 & 1.30 (0.87) & 0.87 (0.75) & 1.03 (0.84) & 0.99 \\ %
2.2 & 6 & 0.7 & 1.35 (1.04) & 0.91 (0.89) & 1.07 (1.01) & 1.19 \\ %
\hline %
\end{tabular} %
\end{center} %
\caption{The same values as in Table~\ref{tableUnshifted} but
calculated using the EBL model of Ref.~\cite{Kneiske:2003tx} } %
\label{tableUnshiftedKneiskeBFebl} %
\end{table*} %

Table~\ref{tableUnshifted} clearly shows that IGRB measurements
provide a significant constraint for the UHECR models with protons as
primaries. In the case of minimal EBL~\cite{Inoue:2012bk} only models
with soft enough injection spectra, $\gamma_g \gtrsim 2.6$, and weak
evolution $m \leq 1$ survive the IGRB constraint. This tension may be
avoided in the case of galactic foreground model B, which predicts a
lower $\gamma$-ray flux.

\begin{figure*}[ht!] 
\begin{center}
 \begin{minipage}[ht]{80mm}
 \centering
 \vspace{-2mm}
 \includegraphics[width=80 mm, height=55.5 mm]{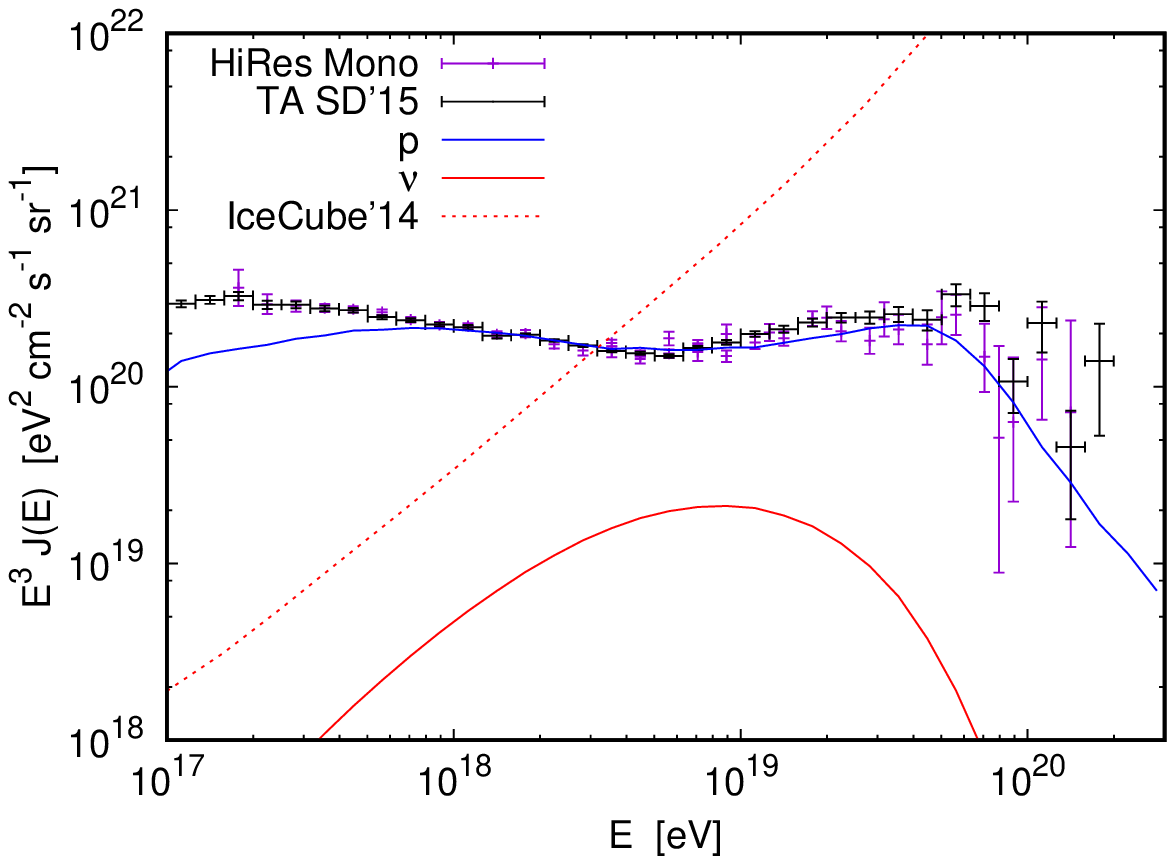}
 \subcaption{UHECR and secondary $\nu$} %
 \end{minipage}
 \hspace{3mm}
 \begin{minipage}[h]{80mm}
 \centering
 \includegraphics[width=80 mm, height=55.5 mm]{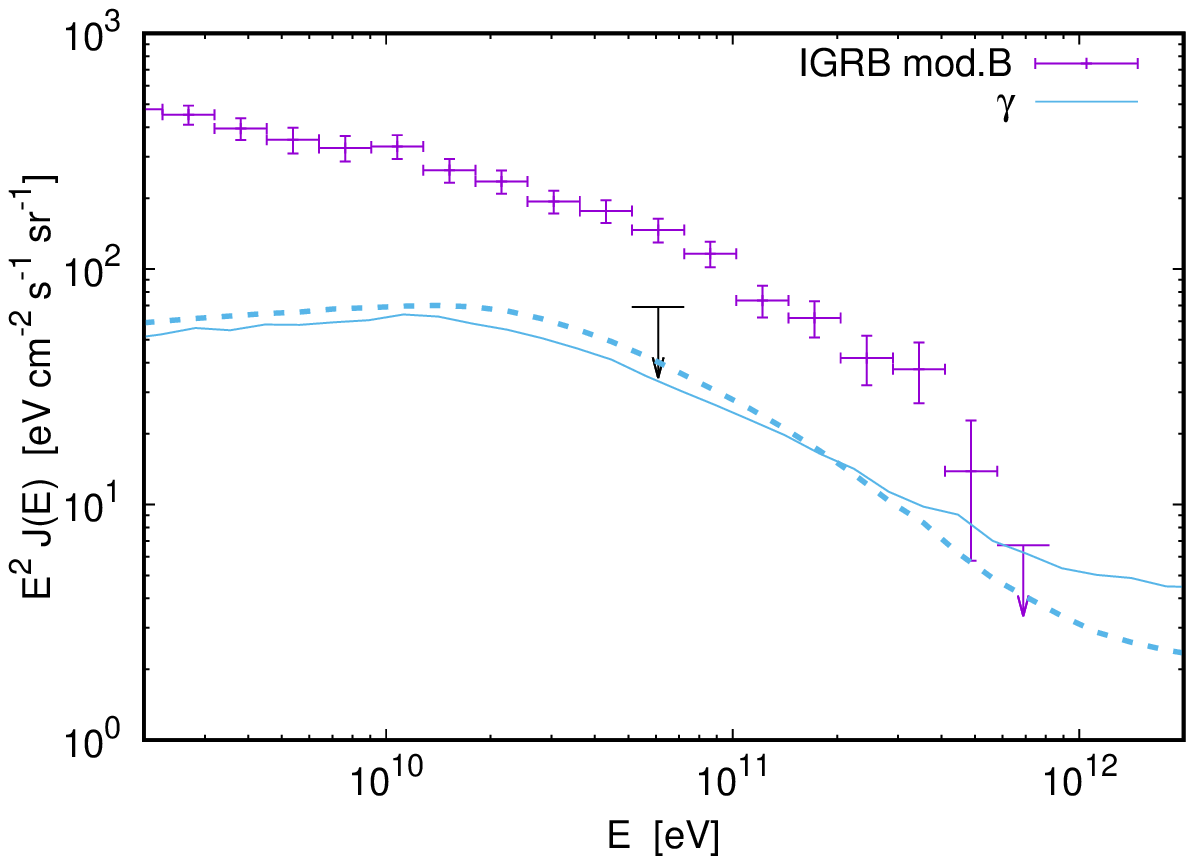}
 \subcaption{secondary $\gamma$} %
 \end{minipage}
 \end{center}
\vspace{-4 mm}%
\caption{
Energy spectra of protons and neutrinos (left panel) and of
cascade photons (right panel) from sources emitting protons with
$\gamma_g=2.6$, $m=1$ and $z_{\rm max}=5$ normalized on TA
spectrum~\cite{TheTelescopeArray:2015mgw}. Also, the Fermi IGRB
measurements are shown for galactic foreground model B, as well as
secondary $\nu$-spectrum along with IceCube neutrino 'differential 
flux' upper limit~\cite{Ishihara:2015xuv}. The Fermi LAT constraint 
of Eq.~(\ref{ratio50GeV}) is shown by the black arrow. EBL models of
Ref.~\cite{Inoue:2012bk} (solid lines) and~\cite{Kneiske:2003tx}
(dashed line) were used in calculations. Only $\gamma$-ray spectrum
is shown for EBL model~\cite{Kneiske:2003tx} since $p$- and
$\nu$-spectra calculated using different EBL models are practically
indistinguishable.
}
\label{specUnshiftedOK}
\end{figure*} 

With EBL of Ref.~\cite{Kneiske:2003tx} (see Table
\ref{tableUnshiftedKneiskeBFebl}) more models survive the $\gamma$-ray
constraint, though SFR evolution model is still excluded.

It is interesting to note that the results presented in columns 5 and
6 of Table~\ref{tableUnshiftedKneiskeBFebl} show the agreement of
UHECR models with Fermi LAT flux in the case of galactic subtractions
B or C and $\gamma_g\geq2.5$. Moreover, the condition
(\ref{ratioIGRB}) is fulfilled even for harder injection spectra while
condition (\ref{ratio50GeV}) in this case  may be satisfied only by
constraining maximal source redshift $z_{\rm max}\lesssim0.7$.

In Fig.~\ref{specUnshiftedOK} we show the spectra of cosmic rays and
secondaries (high energy $\gamma$'s and $\nu$'s) from their
interactions with CMB and EBL in the model with $\gamma_g=2.6$ and
$m=1$. The upper limit at 50~GeV shown in Fig.~\ref{specUnshiftedOK}b
is the constraint of Eq.~(\ref{ratio50GeV}) due to unresolved sources
{\cite{DiMauro:2016cbj}.

We see that in the case of minimal EBL flux of cascade photons is very
close to the Fermi LAT upper bound in the highest energy IGRB bin
assuming galactic foreground model B. Note that for A and C models,
which predict higher contributions of galactic foreground and therefore
lower IGRB flux, the resulting cascade photon flux produced by UHE
protons exceeds the IGRB flux in the last energy bin.

In Fig.~\ref{specUnshiftedNotOK} we show the model with $\gamma_g =
2.4$ and SFR evolution which almost saturates the allowed flux of the
cascade photons for the EBL model of Ref.~\cite{Inoue:2012bk}. %
Summarizing results presented in Tables \ref{tableUnshifted} and
\ref{tableUnshiftedKneiskeBFebl} first of all one notes that the
strongest constraints are provided by HEB in the Fermi LAT spectra.
The constraints depend strongly on the galactic subtraction model A,
B, and C used in the Fermi LAT analysis, and on model of the EBL (we
use two models Refs.~\cite{Inoue:2012bk} and \cite{Kneiske:2003tx}).
For exclusion of each UHECR model one must choose EBL providing the
lowest calculated flux and the highest Fermi LAT flux. To be
conservative about unresolved source Fermi LAT flux we consider both
Eq.~(\ref{ratioIGRB}) and Eq.~(\ref{ratio50GeV}).

From Tables \ref{tableUnshifted} and \ref{tableUnshiftedKneiskeBFebl}
we see that some pure proton composition models with $\gamma_g\leq2.5$
fitting entire CR spectrum are excluded by Fermi LAT fluxes.

One way to relax this tension is to shift the experimental data energy
scale downwards by an amount allowed by systematic errors. In all
existing UHECR experiments, the systematic errors in energy
determination are quite large. In the case of TA data, these errors
are $\sim 20\%$~\cite{TheTelescopeArray:2015mgw}. We estimate the
effect of systematic errors by fitting TA SD spectrum with energy
scale shifted by 20\% towards lower energies. It is naturally expected
that $E^{-\gamma}$ spectrum shifted downwards produces fewer cascade
photons because of diminishing the number of protons at the threshold
of $e^+e^-$ pair-production. The results of our calculations are
presented in Tables~\ref{tableShifted} and
\ref{tableShiftedKneiskeBFebl}. One can see that more models indeed
become acceptable in this case, in particular those with SFR evolution
for both galactic foregrounds B and C. The comparison of this model
with Fermi LAT and IceCube data is shown in Fig.~\ref{specShiftedOK}.

\begin{table*}[ht] %
\begin{center} %
\begin{tabular}{c|c|c|c|c|c|c} \hline %
			$\gamma_g$  &  m  &  $z_{\rm max}$  &$\eta_{\rm\gamma}$
			($\tilde{\eta}_{\rm\gamma}$) [A] & $\eta_{\rm\gamma}$
			($\tilde{\eta}_{\rm\gamma}$) [B] & $\eta_{\rm\gamma}$
			($\tilde{\eta}_{\rm\gamma}$) [C] & $\bar{N_{\rm\nu}}$ \\ \hline			
			2.6 & 0 & 5 & 0.80 (0.26) & 0.53 (0.23) & 0.63 (0.26) & 0.26 \\
			2.6 & 0 & 1 & 0.79 (0.23) & 0.53 (0.20) & 0.63 (0.22) & 0.15 \\
			2.5 & 2 & 5 & 1.00 (0.54) & 0.67 (0.46) & 0.79 (0.52) & 1.40 \\
			2.5 & 2 & 1 & 0.98 (0.37) & 0.66 (0.32) & 0.78 (0.36) & 0.30 \\
			2.4 & SFR & 5 & 1.18 (0.76) & 0.79 (0.65) & 0.94 (0.73) & 1.43 \\
			2.4 & 3 & 5 & 1.16 (0.87) & 0.77 (0.75) & 0.92 (0.84) & 5.00 \\
			2.3 & 4 & 1 & 1.29 (0.67) & 0.86 (0.57) & 1.02 (0.64) & 0.81 \\
			2.2 & 5 & 1 & 1.47 (0.90) & 0.98 (0.77) & 1.16 (0.87) & 1.34 \\
			2.2 & 5 & 0.7 & 1.38 (0.53) & 0.92 (0.46) & 1.09 (0.51) & 0.64 \\
			2.2 & 6 & 0.7 & 1.46 (0.62) & 0.98 (0.53) & 1.15 (0.60) & 0.75 \\
\hline %
\end{tabular} %
\end{center} %
\caption{
	Maximal ratios $\eta_{\rm \gamma}$,
$\tilde{\eta}_{\rm\gamma}$ (assuming galactic $\gamma$-foreground
models A, B or C) for several representative proton
source models fitting TA spectrum with energy scale downshifted by
20\%. The ratios higher than 1 are in conflict with secondary Fermi LAT IGRB flux
measurements. Also shown the expectation value of the neutrino events $\bar{N_{\rm\nu}}$ with energy $E_\nu>10$~PeV assuming IceCube 7 year exposure from Fig.1 of Ref.~\cite{Aartsen:2016ngq}. Models with $\bar{N_{\rm\nu}}>2.3$ have Poisson probability less than 10\%. The spectra are calculated using EBL model of
Ref.~\cite{Inoue:2012bk}.}
\label{tableShifted} %
\end{table*} %
\begin{table*}[ht] %
\begin{center} %
\begin{tabular}{c|c|c|c|c|c|c} \hline %
$\gamma_g$  &  m  &  $z_{\rm max}$  &$\eta_{\rm\gamma}$
($\tilde{\eta}_{\rm\gamma}$) [A] & $\eta_{\rm\gamma}$
($\tilde{\eta}_{\rm\gamma}$) [B] & $\eta_{\rm\gamma}$
($\tilde{\eta}_{\rm\gamma}$) [C] & $\bar{N_{\rm\nu}}$ \\ \hline %
2.6 & 0 & 5 & 0.53 (0.29) & 0.36 (0.25) & 0.42 (0.28) & 0.26 \\ %
2.6 & 0 & 1 & 0.53 (0.24) & 0.35 (0.20) & 0.42 (0.23) & 0.15 \\ %
2.5 & 2 & 5 & 0.64 (0.65) & 0.43 (0.56) & 0.51 (0.63) & 1.40 \\ %
2.5 & 2 & 1 & 0.62 (0.39) & 0.42 (0.34) & 0.49 (0.38) & 0.30 \\ %
2.4 & SFR & 5 & 0.73 (0.84) & 0.49 (0.72) & 0.58 (0.82) & 1.43 \\ %
2.4 & 3 & 5 & 1.27 (1.13) & 0.87 (0.97) & 1.12 (1.09) & 5.00 \\ %
2.3 & 4 & 1 & 0.77 (0.71) & 0.52 (0.61) & 0.61 (0.68) & 0.81 \\ %
2.2 & 5 & 1 & 0.86 (0.96) & 0.57 (0.82) & 0.68 (0.93) & 1.34 \\ %
2.2 & 5 & 0.7 & 0.83 (0.56) & 0.56 (0.48) & 0.66 (0.54) & 0.64 \\
2.2 & 6 & 0.7 & 0.85 (0.66) & 0.57 (0.56) & 0.67 (0.63) & 0.75 \\
\hline %
\end{tabular} %
\end{center} %
\caption{Same as Table~\ref{tableShifted} but calculated using EBL
model of Ref.~\cite{Kneiske:2003tx} } %
\label{tableShiftedKneiskeBFebl} %
\end{table*} %

\begin{figure}[ht!] 
\begin{center} %
\includegraphics[width=80mm]{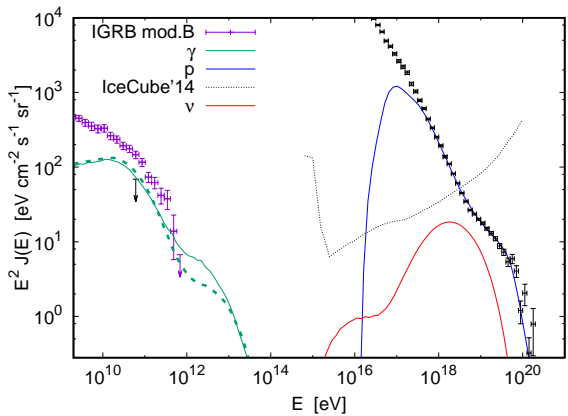} %
\end{center} %
\caption{Energy spectrum of cosmic rays and secondary $\nu$'s and
$\gamma$'s from sources emitting protons with $\gamma_g=2.4$ and
evolution corresponding to star formation rate
(SFR) \cite{Yuksel:2008cu} normalized on TA spectrum
\cite{TheTelescopeArray:2015mgw}. 
The Fermi LAT constraint given by Eq.~(\ref{ratio50GeV}) is shown
by the black arrow. The EBL models of Refs.~\cite{Inoue:2012bk} (solid
lines) and \cite{Kneiske:2003tx} (dashed line) are used in
calculations. The $\gamma$-ray spectrum is shown only for EBL of
model Ref.~\cite{Kneiske:2003tx} since $p-$ and $\nu$-spectra calculated
using different EBL models are almost indistinguishable.} %
\label{specUnshiftedNotOK} %
\end{figure} %

\begin{figure}[ht] 
\begin{center} %
\includegraphics[width=80mm]{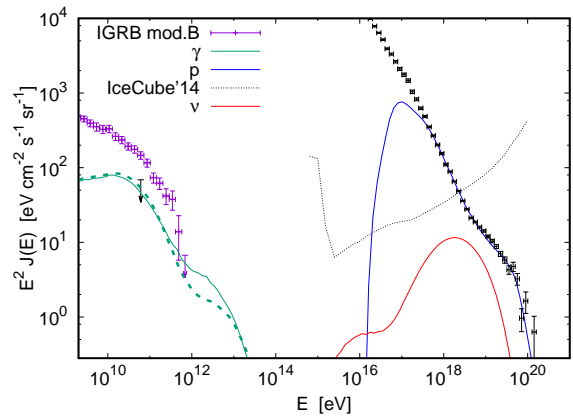} %
\end{center} %
\caption{Same as Fig.~\ref{specUnshiftedNotOK} but for UHECR spectrum
normalized by the TA spectrum with energy scale downshifted by 20\%.} %
\label{specShiftedOK} %
\end{figure} %

\subsection{$(1-4)$~EeV band} %
\label{sec:1-4EeV} %
All biggest detectors agree that mass composition in the energy range
$(1-4)$ EeV is light. The consistency of this result can be tested
independently by calculation of the secondary cascade radiation. In
the recent work Ref.~\cite{Liu:2016brs} authors found a contradiction
between the assumption of pure proton composition in the discussed
energy range and the Fermi LAT IGRB data.

\begin{figure*}[ht] 
\begin{center} %
\begin{minipage}[ht]{58 mm} %
\centering %
\includegraphics[width=58 mm, height=48.5 mm]{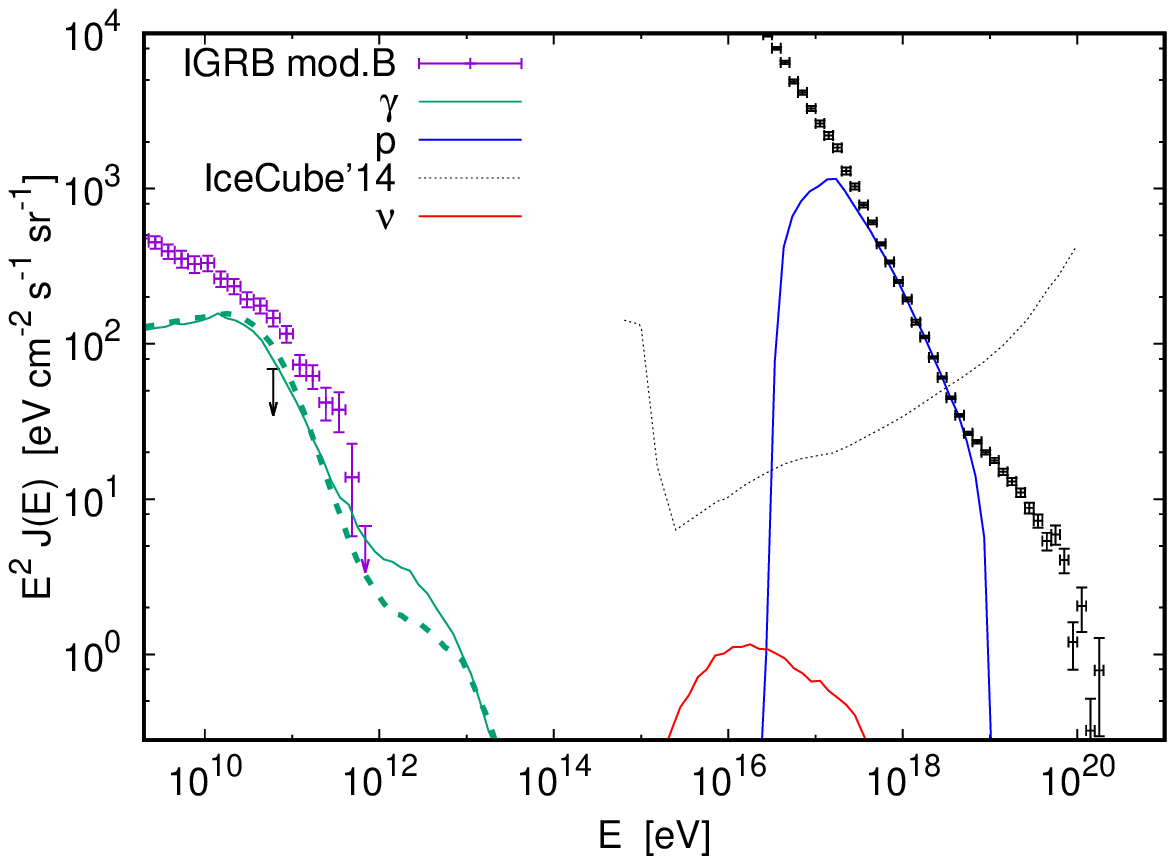}
\subcaption{$\gamma_g=2.1$, $m=3.9$, $z_{\rm max}=2$} %
\end{minipage} %
~
\begin{minipage}[h]{58 mm} %
\centering %
\includegraphics[width=58 mm, height=48.5 mm]{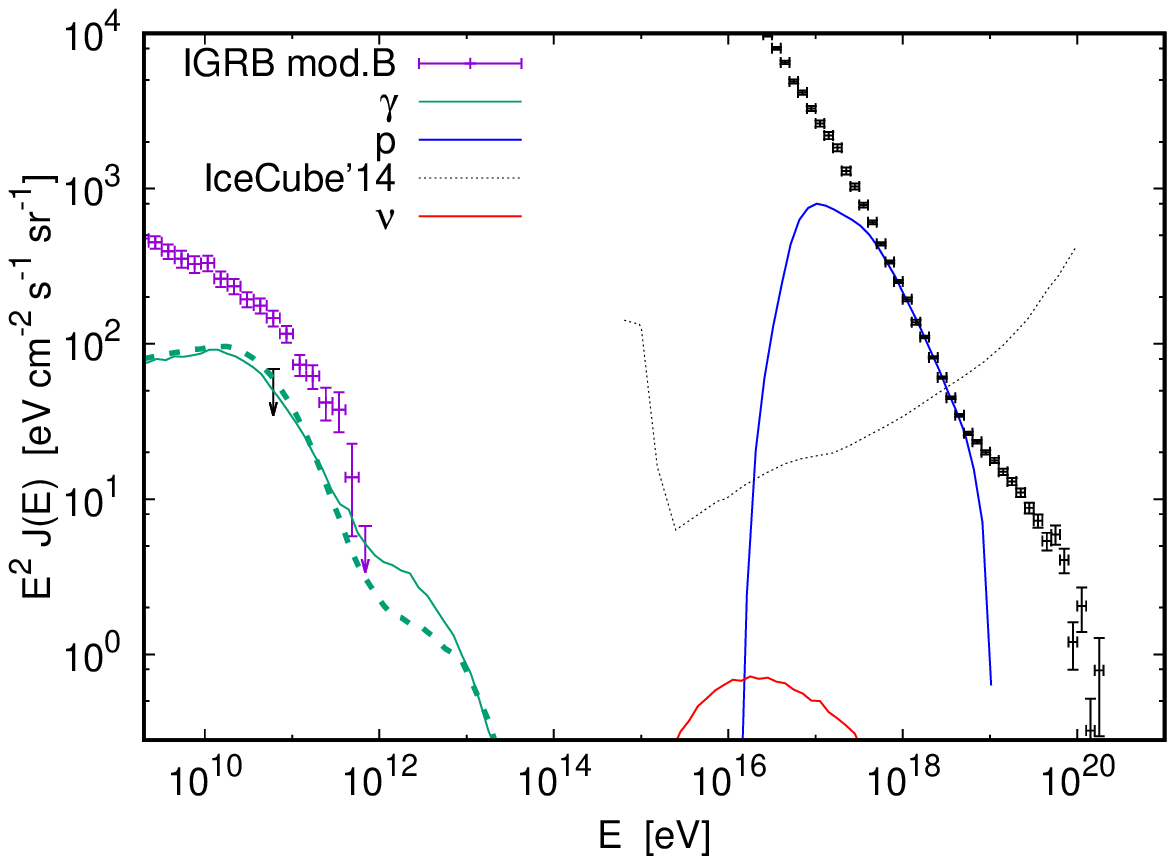}
\subcaption{$\gamma_g=2.19$, SFR evolution} %
\end{minipage} %
~
\begin{minipage}[h]{58 mm} %
\centering %
\includegraphics[width=58 mm, height=48.5 mm]{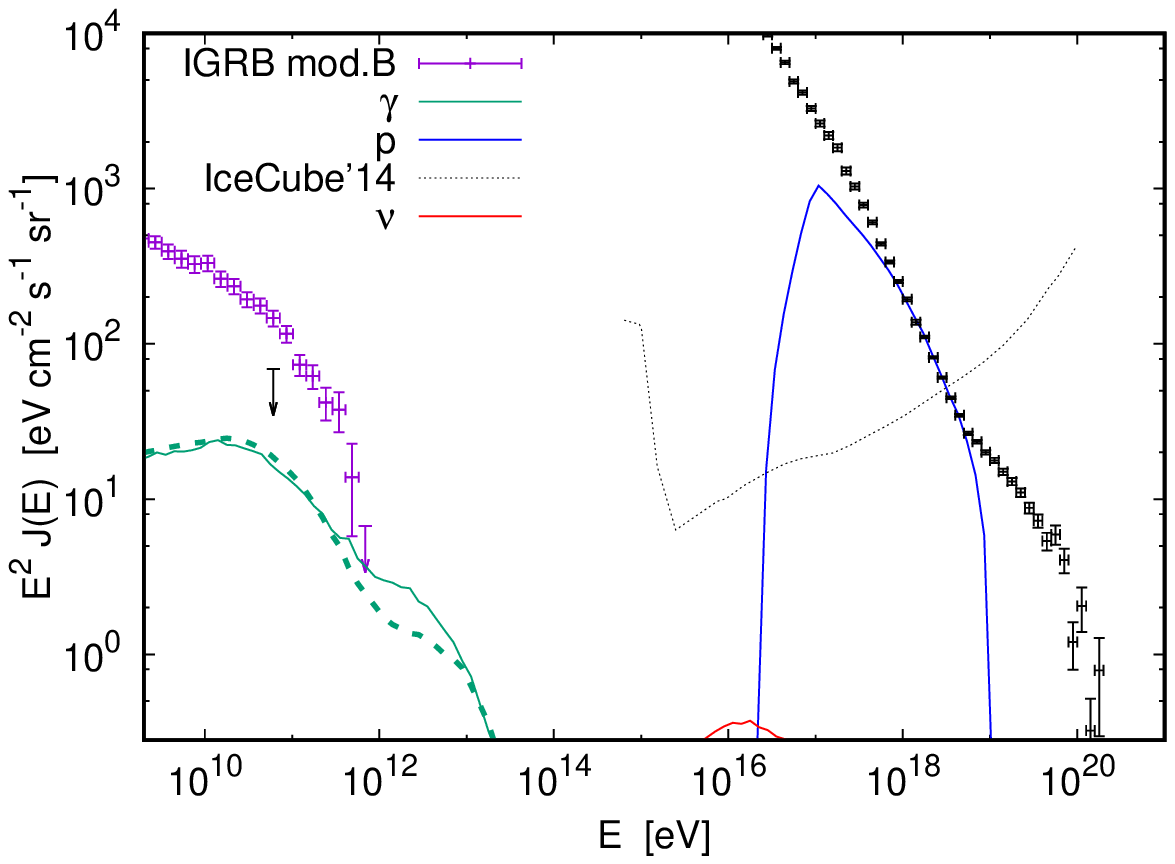}
\subcaption{$\gamma_g=2.6$, $m=0$, $z_{\rm max}=3$} %
\end{minipage} %
\end{center} %
\caption{Energy spectra of cosmic rays and secondary $\nu$ and 
$\gamma$ from proton sources with $E_{\max} = 10$~EeV and fitting the
TA spectrum~\cite{TheTelescopeArray:2015mgw} in the energy range $(1 -
4)$~EeV. The Fermi LAT constraint of Eq.~(\ref{ratio50GeV}) is shown
by the black arrow. Cascade $\gamma$-fluxes calculated using EBL
models of Ref.~\cite{Inoue:2012bk} and Ref.~\cite{Kneiske:2003tx} are
shown by solid and dashed lines, respectively. The cosmic ray and
neutrino spectra calculated using either EBL model are almost
indistinguishable.
} %
\label{spec1e19} %
\end{figure*} %

In fact, $\gamma$-ray production in the discussed energy range is
related to the generation rate of protons at early times, i.e.\ at
large redshifts $z$, and thus it is model dependent. In this
subsection we demonstrate that there are  models in which the
$\gamma$-ray radiation from the existing $(1 - 4)$ EeV proton band,
and from production by protons at larger $z$, does not exceed  the
Fermi LAT limit. In our calculations we construct the models fitting
total cosmic ray flux just in the $(1 - 4)$ EeV energy band, and
predicting the flux which is less than the observed one outside this
energy range. For this we introduce artificial cutoffs in the
generation rate at energies above and below $(1 - 4)$ EeV (see
Fig.~\ref{spec1e19}).

In calculations below we use only two additional assumptions: the
Fermi LAT galactic component is that of model $B$, and the EBL is one
of two models, Ref.~\cite{Inoue:2012bk} or Ref.~\cite{Kneiske:2003tx}.
We made the calculations for the two extreme values of generation
indexes $\gamma_g = 2.6$ and $\gamma_g = 2.1$ as well as for the 
representative case of SFR evolution. The calculated $\gamma$-flux
together with Fermi LAT upper limits are shown in
Fig.~\ref{spec1e19}. One may see that in this case fewer models are
constrained by Fermi LAT. In particular, SFR evolving sources are not
excluded even without shifting experimental energy scale. This is an
expected result since by cutting injection at $E_{\rm max} = 10$ EeV
we have also decreased the contribution to EM component.
Interestingly, the model with hard injection spectrum $\gamma_g = 2.1$
and evolution stronger than of SFR, shown in Fig.~\ref{spec1e19}a, is
prohibited only by constraint Eq.~(\ref{ratio50GeV}) and formally
satisfies the Fermi LAT IGRB bound.

Thus we conclude that pure proton contents of the observed $(1 - 4)$
EeV energy band produces the $\gamma$-radiation which, at least in some
models, is below the Fermi LAT upper limit.

\subsection{Admixture of nuclei} %
\label{sec:helium} %
Nuclei are less efficient in production of photons and one may think
that if some nuclei are erroneously taken in the experiment as
protons, the calculated $\gamma$-ray production is overestimated as
a prediction. The realistic picture is more complicated.

The most natural case is given by Helium nuclei, which is difficult
to distinguish experimentally from protons.

First, following two papers by Aloisio et al.\ \cite{Aloisio:2008pp,
Aloisio:2010he}, we describe shortly the He$^4$ photo-disintegration
life-time in terms of the Lorentz-factors. The steepening of spectrum
at small Lorentz factor occurs at $\Gamma_c = 4 \times 10^8$ due to
the transition from adiabatic energy losses to photo-disintegration on
EBL. The most noticeable spectrum feature, the Gerasimova-Rozental
cutoff \cite{Gerasimova-Rozental:1961}, occurs at Lorentz factor
$\Gamma_c = 4 \times 10^9$ where the transition from
photo-disintegration on EBL and CMB takes place.

The photo-disintegration of Helium is followed by very fast decays of
the produced secondary nuclei He$^3$, $T$, $D$ and neutron. Hence an
assumption that photo-disintegration of He$^4$ is instantaneously
followed by the production of four protons gives a realistic
description at both spectrum steepenings, $\Gamma_c = 4 \times 10^8$
and $\Gamma_c = 4 \times 10^8$.

The $e^+e^-$ pair-production energy loss plays just a minor role in
the formation of spectrum shape, but this particular process is
responsible for photon production. As was demonstrated in
Ref.~\cite{Aloisio:2008pp, Aloisio:2010he}, the rate of Lorentz-factor
loss $\Gamma^{-1} d\Gamma/dt$ satisfies the following relation for
nuclei $A$ and proton $p$ components
\begin{equation} %
\label{eq:enloss} %
\left( \frac{1}{\Gamma}\frac{d\Gamma}{dt}(\Gamma)\right)_A
 = \frac{Z^2}{A}\left(
\frac{1}{\Gamma}\frac{d\Gamma}{dt} (\Gamma)\right)_p. %
\end{equation} %
Thus the rate of energy loss for He$^4$ with $Z^2/A=1$ is equal to
that of a proton if $\Gamma_A = \Gamma_p$, i.e.\ for nucleus energy $A$
times higher than that of the proton. It means that at equal energies $E_A
= E_p$ the rate of energy loss $(1/E)(dE/dt$) for the nucleus is less than
that for the proton and thus nuclei produce less cascade photons than
protons.

Therefore, in the case of Helium one expects two competing effects in
comparison with protons: i) diminishing of the pair-production energy
loss at the same energy and ii) production of four protons with 4
times lower energy (still active in photon emission) in the prompt
processes of Helium photo-disintegration. The latter process must work
more efficiently for the hard generation spectra.

\begin{figure*}[ht] 
\begin{center} %
\begin{minipage}{80mm} %
\centering %
\includegraphics[width=\textwidth]{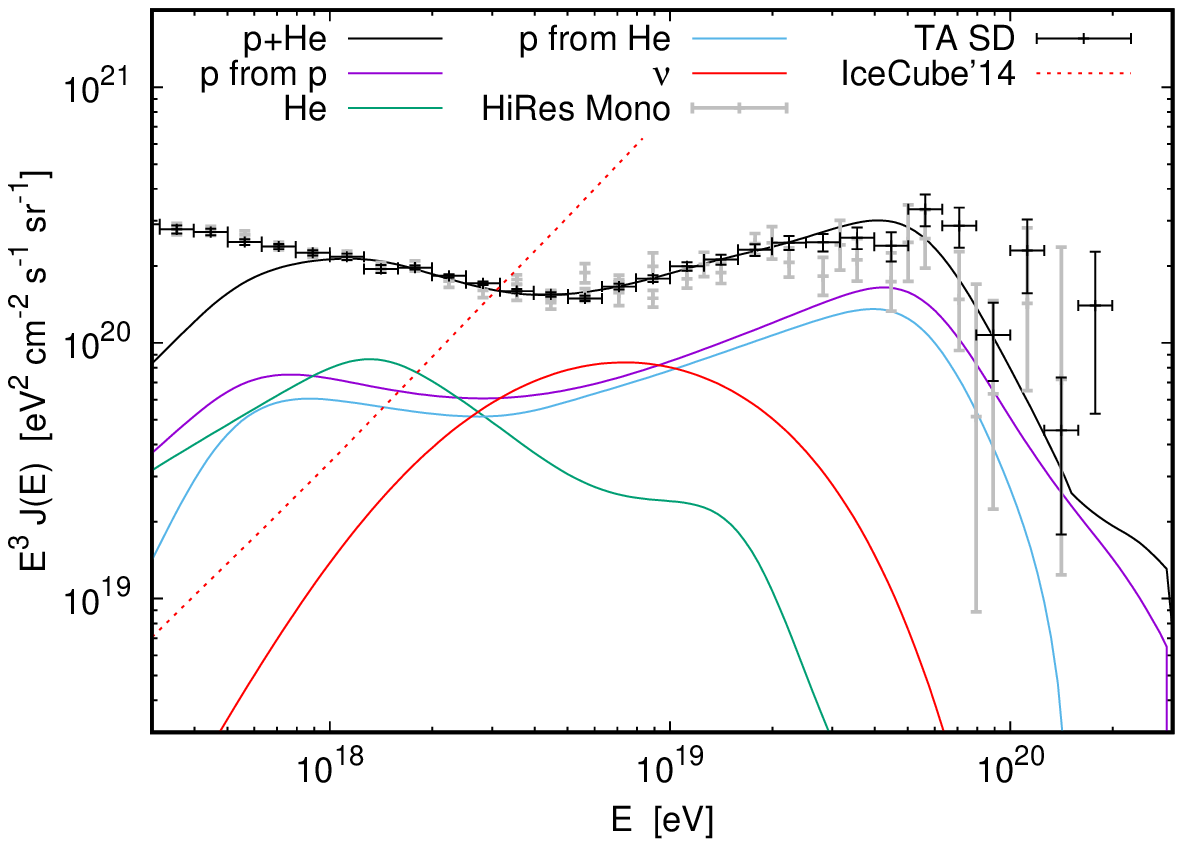} %
\subcaption{Cosmic Rays} %
\label{specUnshiftedHe50:hecr} %
\end{minipage}
~
\begin{minipage}{80mm}
\centering
\includegraphics[width=\textwidth]{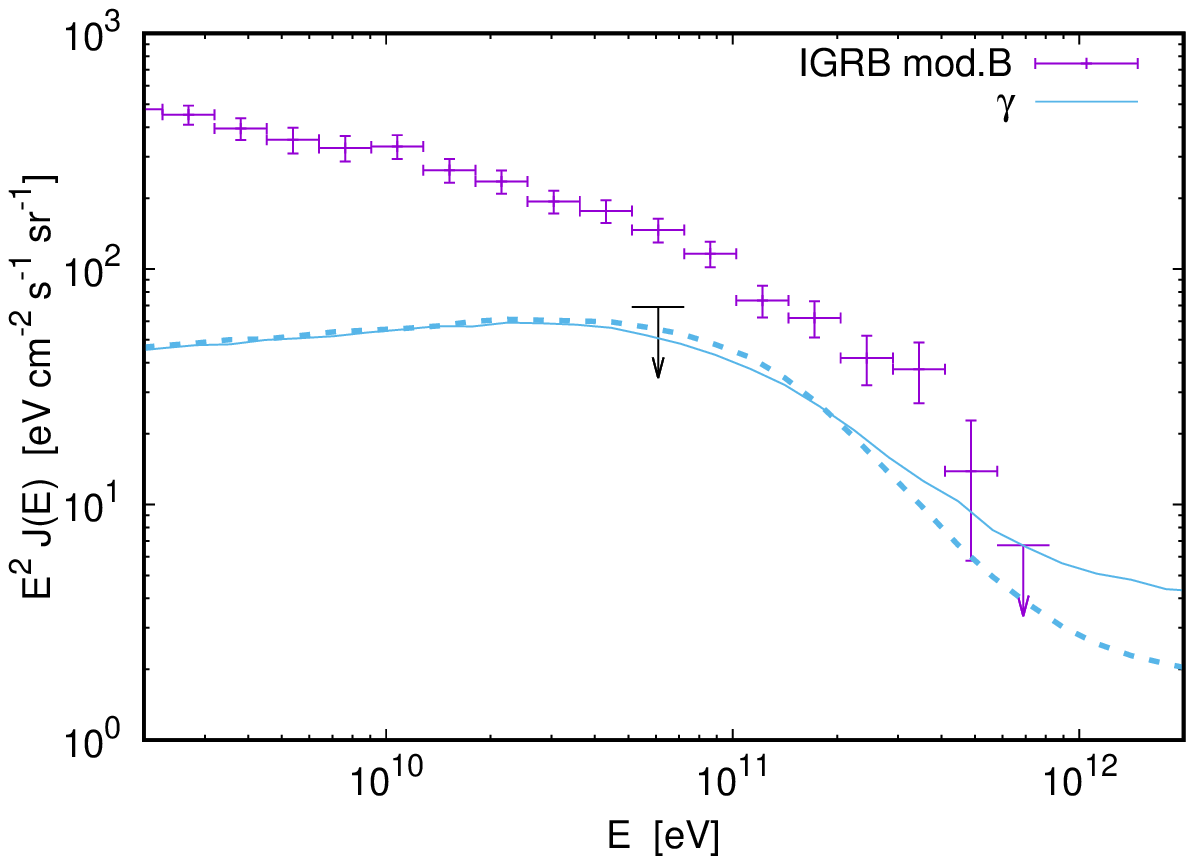}
\subcaption{Secondary $\gamma$'s}
\label{specUnshiftedHe50:gamma}
\end{minipage}
\end{center}
\vspace{-4 mm}%
\caption{Energy spectrum of cosmic rays, secondary neutrinos (left
panel) and cascade photons (right panel) from sources emitting mixture
of protons (48\%) and He (52\%) with $\gamma_g=2.1$, $m=5$ and
$z_{\max}=1$ normalized on TA
spectrum~\cite{TheTelescopeArray:2015mgw}. The constraint of
Eq.~(\ref{ratio50GeV}) is shown by the black arrow (right panel).
$\gamma$-ray spectra are shown for EBL models of
Ref.~\cite{Inoue:2012bk} (solid line) and~\cite{Kneiske:2003tx}
(dashed line). Cosmic ray and $\nu$-spectra are shown only for EBL of
Ref.~\cite{Kneiske:2003tx}
} %
\label{specUnshiftedHe50} %
\end{figure*} %

Below we shall discuss two cases of the mixed composition of protons
and Helium.

The energy spectrum of pure Helium is characterized by the cutoff at
energy $E \sim 1\times 10^{18}$~eV due to photo-disintegration on EBL,
which looks like a maximum in the traditional presentation of spectrum
in the form $E^3 J(E)$. To describe the observed spectrum up to $\sim
100$ EeV one needs another component which in $p+He$ mixing models is
given by protons.

In Fig.~\ref{specUnshiftedHe50} we present the $p+He$ model with
almost extreme mixing which fits the observed spectrum with the ratio
of the production rates $Q_p/Q_{He} = 0.9$. This model is
characterized by generation index $\gamma_g = 2.1$, evolution
parameters $m=5$, $z_{\max}=1$ and maximum of acceleration $E_{\max} =
Z \times 300$~EeV, where $Z$ is the charge. The observed dip in the
energy spectrum is produced by He bump at $E \simeq 1.4$~EeV
superimposed on the proton dip. One can notice that flux of the
secondary protons from photo-disintegration of He$^4$ is practically
equal to the flux of primary protons. The sum of these components
provides the main contribution to the cascade photons, while the
direct contribution from He$^4$ is negligible. The flux of cascade
photons produced in the Fermi HEB in this model is 36\% less than one
predicted by the pure proton source model with the same injection
spectrum.

\begin{figure*}[ht] 
\begin{center} %
\begin{minipage}[b]{80mm} %
\includegraphics[width=\textwidth]{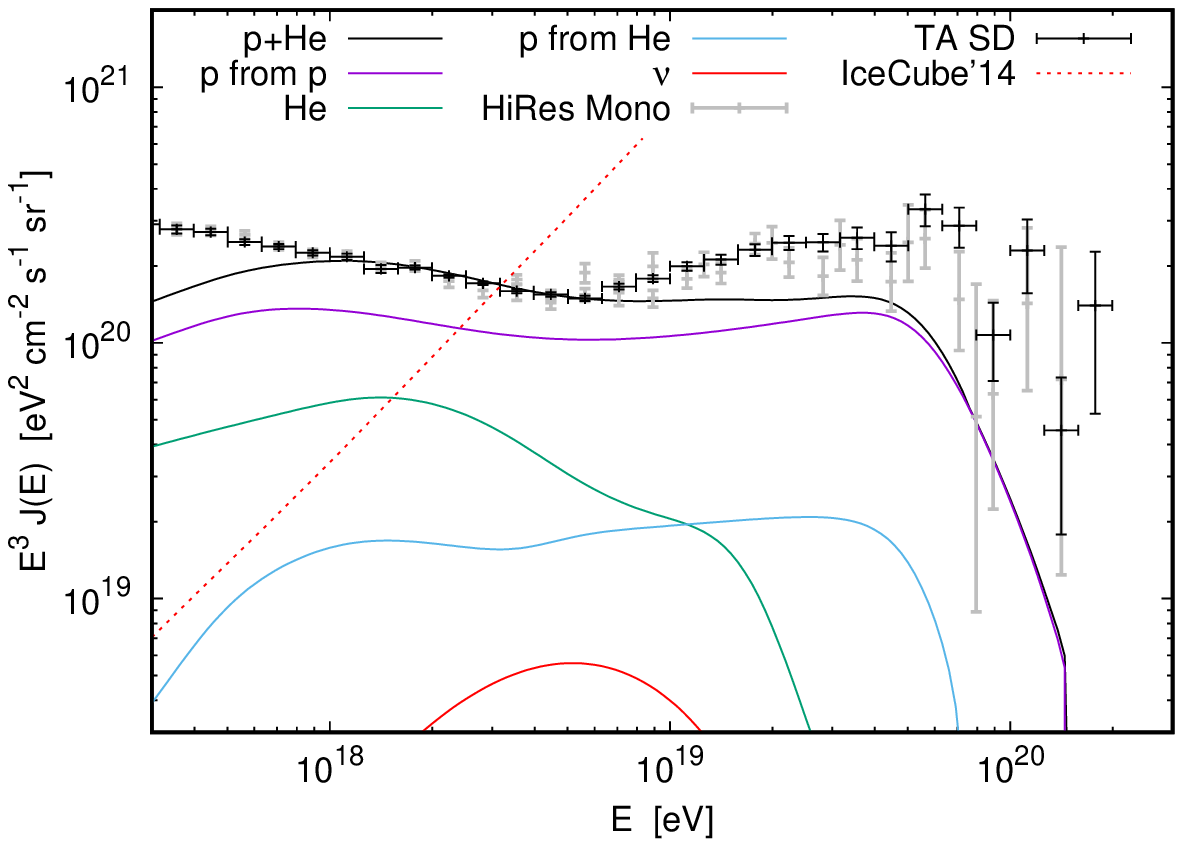} %
\subcaption{Cosmic Rays} %
\label{specUnshiftedHe30:hecr} %
\end{minipage}%
~ %
\begin{minipage}[b]{80mm} %
\includegraphics[width=\textwidth]{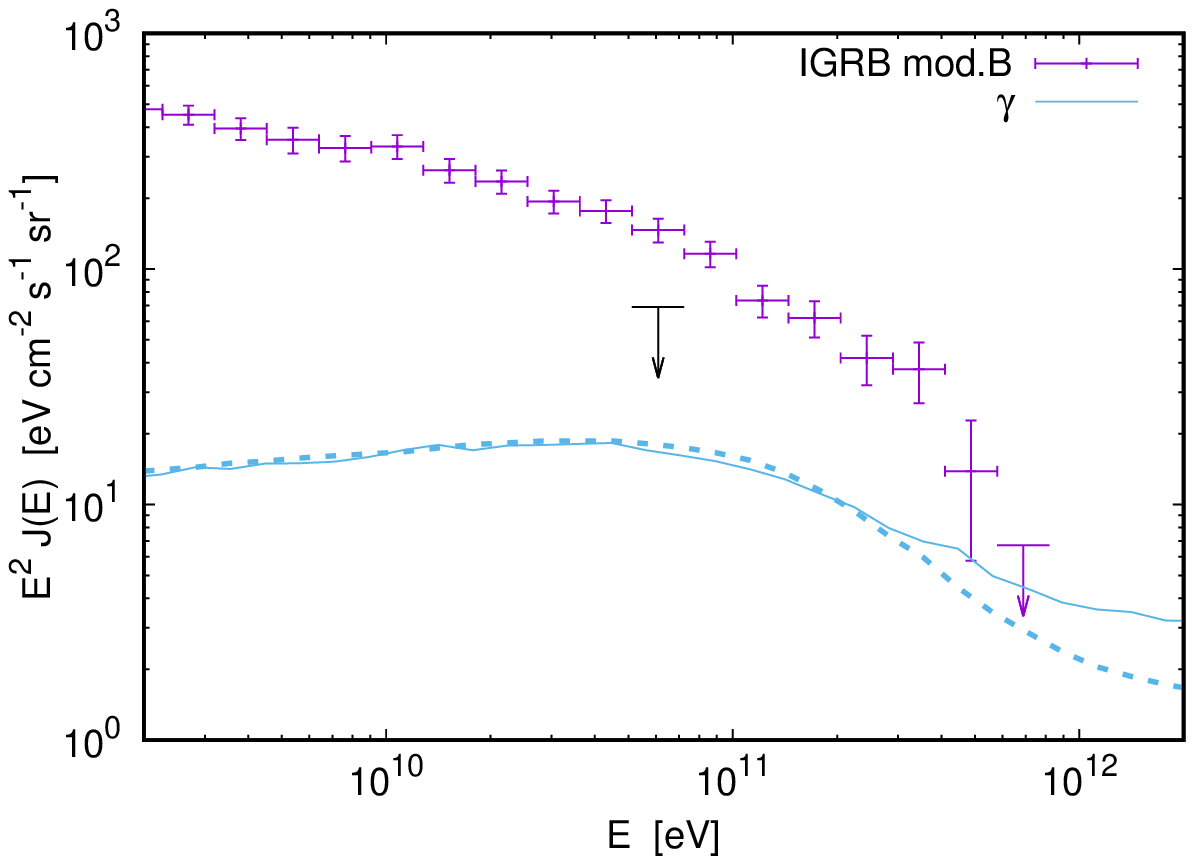} %
\subcaption{Secondary $\gamma$} %
\label{specUnshiftedHe30:gamma} %
\end{minipage} %
\end{center}
\caption{Energy spectrum of cosmic rays, secondary neutrinos (left
panel) and cascade photons (right panel) from sources emitting protons
with 30\% admixture of He with $\gamma_g = 2.6$, $E_{\rm max} =
Z\times 150$  EeV, $m=1$ and $z_{\rm max} = 1$ normalized on TA
spectrum~\cite{TheTelescopeArray:2015mgw}. The Fermi constraint given
by Eq.~(\ref{ratio50GeV}) is shown by the black arrow (right panel).
$\gamma$-ray spectra are shown for EBL models of
Ref.~\cite{Inoue:2012bk} (solid line) and~\cite{Kneiske:2003tx}
(dashed line). Cosmic ray and $\nu$ spectra are only shown for EBL of
Ref.~\cite{Kneiske:2003tx}} %
\label{specUnshiftedHe30} %
\end{figure*} %

In Fig.~\ref{specUnshiftedHe30} we present calculations for soft
proton and He production spectra with $\gamma_g = 2.6$. It is clear
that in this case the production of secondary protons is suppressed;
the direct photon production by Helium is suppressed too for the
general reason discussed above. This model fails to fit the
observational data above 10 EeV, but it allows to suppress the flux of
cascade photons by 26\% compared to pure proton source case. To get a
reasonable fit to the data one should diminish the He content to very
a low level, where suppression of cascade photon flux is approximately
equal to the fraction of He in the generation flux.

\subsection{Sources and magnetic fields} %
\label{sec:sources} %
In case the proton component is the dominant one in UHECR, the
observation of cascade radiation, and in particular the Fermi LAT
observations at present, give information about sources of UHECR. The
photons observed in HEB of Fermi LAT cannot arrive from large
redshifts and thus the sources of UHE protons being parents of photons
from HEB, cannot lie at too large distances. However, the distribution
of protons, the parents of HEB photons, should be wider than that of
photons because of their larger interaction length.

\begin{figure}[ht] 
\begin{center} %
\includegraphics[width=0.45\textwidth]{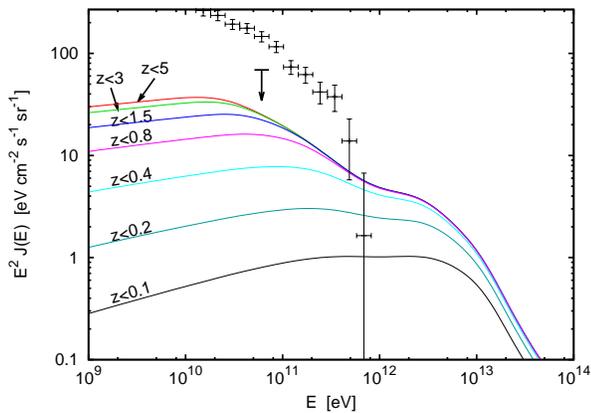} %
\end{center} %
\caption{Cascade photon spectrum from UHE protons with source
injection function $E^{-2.7}$, $m=0$ and $z_{\rm max}=5$ normalized on
TA energy spectrum. Contributions from different redshift ranges of
proton production are shown. Also, the Fermi LAT IGRB measurement
(model B) and constraint of Eq.~(\ref{ratio50GeV}) are shown with
black error bars and arrow respectively.} %
\label{homo_spec} %
\end{figure} %

In Fig.~\ref{homo_spec} the red solid line shows the spectrum of the
cascade photons produced in the typical model of pure proton sources
explaining the observations of TA. In this figure we also show the
contribution of proton sources, located at different redshift ranges,
to the total $\gamma$-ray flux. These fluxes are compared with HEB of
Fermi LAT data~\cite{Ackermann:2014usa}. One may see that cascade flux
in the Fermi LAT HEB $(580 - 820)$~GeV is mostly produced by sources
with redshifts $z < 0.4$, while the cosmologically distant sources,
those with $z > 0.8$, have a weak effect on the last bin.

\begin{figure}[ht] 
\begin{center} %
\includegraphics[width=0.45\textwidth]{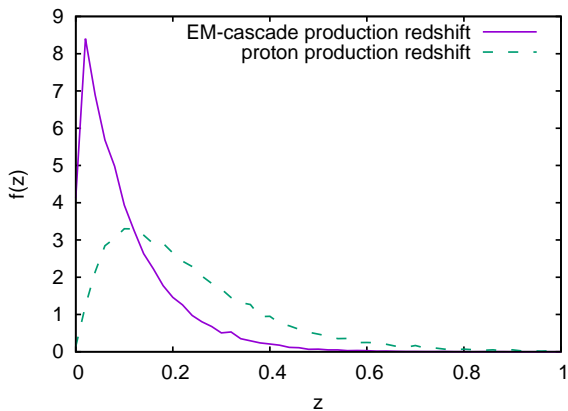} %
\end{center} %
\caption{The distribution of protons (dashed line) and secondary EM
cascade (solid line) production redshifts contributing to the last
Fermi LAT energy bin for the UHECR model used in Fig~\ref{homo_spec}.} %
\label{z_distrib} %
\end{figure} %

Fig.~\ref{z_distrib} presents the redshift distribution of protons and
cascade photons contributing to HEB of the Fermi LAT experiment in the
discussed model. Interestingly, one can see that while the UHE proton
sources may be distant, the EM cascades are initiated relatively
nearby at $z\lesssim 0.1$. This is because photons of HEB are absorbed
at larger distances on EBL radiation, but the parent protons can cross
this distance and produce photons close to the
observer~\cite{Essey:2009ju}. HEB photons are produced nearby: the
maximum of the distribution is located at $z_\gamma^{\max} \simeq
0.01$ with the mean redshift of the distribution $<z_{\gamma}>=0.09$.
The distribution of parent protons production redshift is wider and is
shifted towards higher redshift: $z_p^{\max}=0.1$ and $<z_p>=0.21$.

The results which we have obtained above might in principle depend on
the assumption about the strength of intergalactic magnetic field
(IGMF) which is currently poorly known. Indeed, we assume that
secondary $\gamma$-ray flux from UHECR protons is isotropic in the
entire energy range where IGRB is measured and most importantly in the
last energy bin $E_{\gamma}\lesssim 1$ TeV. However, as it was shown
in Ref.~\cite{Essey:2009ju}, the TeV secondary $\gamma$-rays may point
back to their sources provided that IGMF strength is less than
$10^{-14}$~G. Such events would not contribute to IGRB but to total
EGB flux and therefore one may argue that IGRB bound is irrelevant in
this case. However, the more detailed consideration reveals that just
$6$ resolved sources ($5$ BL Lacs and one of unknown type) with
galactic latitude $|b| > 20^{\circ}$ and redshift $z \leq 0.212$ from
2FHL catalog of Fermi LAT~\cite{Ackermann:2015uya} constitute 100\% of
the resolved source flux in the last energy bin of EGB. At the same
time sources of UHECR are known to be much more numerous. Indeed, the
lack of statistically significant clustering of cosmic rays arrival
directions at small scales leads to a lower limit on the local density
of UHECR sources $n > 10^{-4}$ Mpc$^{-3}$~\cite{Dubovsky:2000gv,
Abreu:2011pf} or roughly $10^5$ sources with $z < 0.21$. This means
that at least in the last energy bin, which is the most important for
us, UHECR sources contribute mostly to IGRB, and EGB bound is
irrelevant for them regardless of the level of IGMF.

\section{Discussion} %
\label{discuss} %
Modern UHECR experiments in which mass composition of UHECR is
measured using atmospheric shower properties such as depth of shower
maximum, result in contradictory conclusions. For this reason the
indirect methods facilitating discrimination of various UHECR
composition models obtain the considerable importance. In this work we
use the method based on calculation of diffuse fluxes of secondary
$\gamma$-rays and neutrinos generated by UHECR during their
propagation. We consider first the models with pure proton composition
which are consistent with TA and HiRes data and then the mixed
composition of protons and Helium with different ratios. These models
include also the dip model which explains the observed feature, the
dip at $(1-40)$ EeV by $e^+e^-$ pair production. We demonstrate that
many proton models are severely constrained by the Fermi LAT IGRB
observations, but there are many models that successfully survive. The
first selection of the proton-dominated models follows from the
condition of describing the TA or HiRes energy spectra; among these
models we choose those with generation indexes $\gamma_g$, $z_{\max}$
and cosmological evolution $(1+z)^m$ up to maximum redshift $z_{\max}$
which do not overproduce the Fermi LAT $\gamma$-ray radiation.

In the case of unshifted TA spectrum only models with $\gamma_g\geq
2.5$ and relatively weak evolution $m \leq 2$ survive. 

The class of surviving models becomes larger when the model B of
galactic $\gamma$-ray radiation is used in the Fermi LAT analysis and
also when the model~\cite{Kneiske:2003tx} is used for EBL. The class
of allowed models becomes further wider with TA energy scale shifted
by 20\% towards lower energies (allowed by systematic errors) the
constraint weakens to $\gamma_g \geq 2.2$ and $m \leq 5$ assuming
$z_{\rm max}=1$ or to $\gamma_g \geq 2.4$ and $m \leq 3$ assuming
$z_{\rm max}=5$ which also includes models with SFR evolution.
Limiting maximal source redshift to a value  $z_{\rm max} \leq 0.7$
allows to include models with $m \leq 6$ and $\gamma_g \geq 2.1$.

Models with strong evolution, which require hard injection spectra and
sufficiently large $z_{\max}$, are constrained also by the neutrino
flux measurements of the IceCube detector~\cite{Aartsen:2016ngq}. However in most cases
modern IGRB constraints on secondary diffuse $\gamma$-ray flux are more
restrictive than the IceCube limit.

All modern experiments show the light nuclei composition in the energy
range $(1-4)$ EeV. Inspired by this observations we consider ad hoc
the proton source models fitting UHECR spectrum only in the
above-indicated energy range. The cascade $\gamma$-ray flux obtained
for such models is also very close to the IGRB constraints. In
particular, the models with $\gamma_g \leq 2.1$ and $m > 3.5$
normalized on TA overproduce cascade photons. However, as we
demonstrated in subsection \ref{sec:helium}, an admixture of Helium
allows to further decrease the cascade $\gamma$-ray flux.

When this paper was in preparation two interesting articles on the
similar subject appeared in arXiv: Refs.~\cite{Gavish:2016tfl,
Liu:2016brs}. Both of them are more pessimistic about proton scenarios
which in our opinion is due to disregarding the uncertainties in the
Fermi LAT data (using the maximal galactic foreground model A) and in
UHECR data (systematic errors), uncertainties in EBL models and
neglecting of the highly possible admixture of He$^4$ nuclei to "pure
proton models", particularly in the $(1 - 4)$~EeV energy band.

We would like to comment on interesting analysis in
Ref.~\cite{Liu:2016brs} concerning the energy range $(1 - 4)$~EeV,
where the authors assume the pure proton mass composition and found
the excess of $\gamma$-radiation over IGRB. The authors argue that
local source overdensity or even galactic sources are required to
avoid contradiction to Fermi data. We think that this problem and its
solution are premature at present. As was demonstrated above, the
contradiction may be avoided by using lower galactic foreground (model
B) in Fermi LAT, or by higher EBL, or by shifting experimental energy
scale within the allowed systematic errors.

The case of pure proton composition in the energy range $(1 - 4)$~EeV,
like in Ref.~\cite{Liu:2016brs}, can be illustrated by
Fig.~\ref{spec1e19} for three values of $\gamma_g = 2.1,\, 2.19$ and
$2.6$, and for different cosmological evolution. In all three cases
when the EBL high-flux model of Ref.~\cite{Kneiske:2003tx} is used
(the dashed lines for calculated $\gamma$-ray spectra in Figs.
\ref{spec1e19}a, \ref{spec1e19}b and \ref{spec1e19}c) the
calculated fluxes are well below the IGRB Fermi LAT upper limit.

Another option not listed above is given by admixture of Helium in the
source spectrum, which can be easily mistaken in observations for
protons. This case is illustrated by Fig.~\ref{specUnshiftedHe30}
where we show the UHECR and secondary $\gamma$-ray spectra in the
source model with $30\%$ Helium admixture. As it was noticed in
Ref.~\cite{Aloisio:2006wv}, the presence of a considerable admixture
of nuclei distorts the shape of the dip. In the figure we fit only the
energy range $(1 -4)$~EeV as in Ref.~\cite{Liu:2016brs}. The cascade
$\gamma$-radiation spectrum in this case is compatible with the Fermi
IGRB bound.

We finally conclude that measurements of the diffuse gamma-radiation
at $E \sim 1$~TeV is a very powerful method to constrain the fraction
of protons in UHECR spectrum. Nowadays, with available statistics and
poor knowledge of the galactic diffuse foreground and EBL, it is
impossible to exclude the pure or almost pure proton composition at
$(1 - 40)$~EeV. However some tension between predictions and
observations of gamma-radiation already exists, especially in the
highest energy bin;  it can be considered as a warning signal and
hence a motivation for consideration of alternative solutions.

The discussed problem will be one of the important tasks for the
future CTA~\cite{Consortium:2010bc}.

\section*{Acknowledgments}
Work of OK was supported by the
Russian Science Foundation, grant 14-12-01340. OK is grateful to GSSI
and LNGS for hospitality. %

\bibliographystyle{elsarticle-num}
\bibliography{last-bin}
\end{document}